\documentclass[iop,letterpaper]{emulateapj}
\usepackage{times}
\usepackage{courier}
\usepackage{chngpage}
\usepackage{color}
\usepackage{float}
\usepackage[hidelinks]{hyperref}
\usepackage[fleqn]{amsmath}
\usepackage{relsize}
\usepackage{amssymb}
\usepackage{multirow}

\slugcomment{Accepted 2015-May-6 to ApJ}

\shorttitle{Asteroids in {\it GALEX}}
\shortauthors{Waszczak et al.}

\begin{document}

\title{Asteroids in {\it GALEX}: Near-ultraviolet photometry of the major taxonomic groups}

\author{Adam Waszczak$^1$, Eran O. Ofek$^{2,3}$, Shrinivas R. Kulkarni$^1$}

\affil{\\$^1$Division of Geological and Planetary Sciences, California Institute of Technology, Pasadena, CA 91125, USA: \href{mailto:waszczak@caltech.edu}{\color{blue}{waszczak@caltech.edu}}\\
         $^2$Benoziyo Center for Astrophysics, Faculty of Physics, Weizmann Institute of Science, Rehovot 76100, Israel\\
         $^3$Helen Kimmel Center for Planetary Science, Weizmann Institute of Science, Rehovot 76100, Israel\\}

\begin{abstract}
We present ultraviolet photometry (NUV band, 180--280 nm) of 405 asteroids observed serendipitously by the {\it Galaxy Evolution Explorer} ({\it GALEX}) from 2003--2012. All asteroids in this sample were detected by {\it GALEX} at least twice. Unambiguous visible-color-based taxonomic labels (C type versus S type) exist for 315 of these asteroids; of these, thermal-infrared-based diameters are available for 245. We derive $\text{NUV}-V$ color using two independent models to predict the visual magnitude $V$ at each NUV-detection epoch. Both $V$ models produce $\text{NUV}-V$ distributions in which the S types are redder than C types with more than 8$\sigma$ confidence. This confirms that the S types' redder spectral slopes in the visible remain redder than the C types' into the NUV, this redness being consistent with absorption by silica-containing rocks. The \emph{GALEX} asteroid data confirm earlier results from the {\it International Ultraviolet Explorer}, which two decades ago produced the only other sizeable set of UV asteroid photometry. The \emph{GALEX}-derived $\text{NUV}-V$ data also agree with previously published \emph{Hubble Space Telescope} (\emph{HST}) UV observations of asteroids 21 Lutetia and 1 Ceres. Both the \emph{HST} and \emph{GALEX} data indicate that NUV band is less useful than $u$ band for distinguishing subgroups within the greater population of visible-color-defined C types (notably, M types and G types).

\end{abstract}

\keywords{surveys --- minor planets, asteroids: general --- solar system: general}

\section{Introduction}

As in visible wavelengths, ultraviolet flux from asteroids is reflected sunlight. However, the steep drop in the solar spectrum shortward of $\sim$300 nm (Figure 1) makes asteroids orders of magnitude fainter in the UV than in the visible. For this reason---as well as the strong UV absorption by atmospheric ozone---UV observations of asteroids typically employ the {\it Hubble Space Telescope} ({\it HST}) or specialized instruments on a space-mission payload physically closer to the asteroid. These constraints have generally prohibited large-sample demographic studies of asteroids in the UV.

Predating \emph{HST}, the {\it International Ultraviolet Explorer} ({\it IUE}) targeted 45 asteroids from 1978--1992, producing what remains to date the largest published sample of near-UV asteroid spectra \citep{roe94}, specifically in the range of 230--325 nm. The {\it IUE} data show evidence of clustering, principally with respect to geometric UV albedo. This clustering becomes further evident when coarsely-defined visible spectral type is included as a categorical parameter for each object (C, S and M types being the classes considered in the original work). Comparing the {\it IUE}-derived geometric UV albedos for each class with the geometric visible albedos demonstrated that the S types, which are redder-colored in the visible (specifically, 400--800 nm), remain redder than C types into the NUV.

The \emph{IUE} data, combined with previously-measured visible spectra, suggested that asteroid reflectances over the entire near-UV to visible wavelength range (200--800 nm) are generally consistent with those of silica-bearing rocks ({\it e.g.}, \citealp{wag87}). To first order, this trend is characterized by an increase in a rock's reflectance at longer wavelengths, generally attributable to the decreased fraction of volume-scattered light, {\it i.e.}, light which penetrates into the mineral grains. The intensity of volume-scattered light varies as $\exp(-kd/\lambda)$, where $d$ is the grain size, $k$ is the imaginary part of the index of refraction, and $\lambda$ is the wavelength. Once refracted into the grains, volume-scattered light is subject to absorption by various ($\lambda$-dependent) interactions with the mineral's crystalline structure. At sufficiently short wavelengths ({\it i.e.}, well into the ultraviolet region), most incident light penetrates the grains and is absorbed, while the small amount of measured reflected light is predominantly scattered directly from the grain surface. Precise characterization of this transition between surface-dominated and surface plus volume-scattered reflectance---as well as the identification of any additional mineral-specific spectral features---is therefore useful for tying astronomical observations of asteroids to laboratory-measured analogs, including lunar and meteoritic samples.

\begin{table*}
\centering
\caption{Observations of asteroids detected in {\it GALEX} NUV images. Includes 1,342 detections of 405 asteroids detected at least twice. }
\begin{tabular}{cccccccccc}
\hline\\[-2ex]
asteroid & observation & detected   &  detected   & position        & NUV & NUV mag    & MPC-predicted & exposure & unique database ID\\
 number  & date (UT)   & R.A. (deg) &  Dec. (deg) & residual ($''$) & mag & uncertainty& visible mag ($V_\text{MPC}$) & time (s) &(\texttt{objID} key in CasJobs)\\[1ex]
\hline
    1 & 2011-10-12.65797 & 355.22180 & -18.47266 & 0.5 & 14.38 & 0.01 &  8.0 &   91 & 6380556162844065792 \\
    1 & 2011-10-21.42095 & 353.95617 & -18.39556 & 0.2 & 14.67 & 0.01 &  8.1 &   80 & 6380556163951362048 \\
    3 & 2005-12-26.72536 &  75.12144 &  -1.31299 & 1.4 & 14.43 & 0.01 &  7.8 &   80 & 6381858059773280256 \\
    3 & 2011-04-17.04995 & 166.80140 &   8.52138 & 1.4 & 16.30 & 0.01 &  9.8 & 1513 & 3855329770719936512 \\
    6 & 2006-08-29.56208 & 309.25038 & -19.31464 & 0.9 & 14.78 & 0.01 &  8.4 &  112 & 6379782093773209600 \\
    6 & 2005-05-07.46079 & 204.16985 &  11.58510 & 0.4 & 16.67 & 0.02 & 10.2 &  112 & 6378656257217134592 \\
    8 & 2004-12-21.41462 & 122.52225 &  19.17684 & 0.7 & 15.63 & 0.02 &  9.0 &   92 & 6377776615736213504 \\
    8 & 2004-12-21.48313 & 122.51009 &  19.18345 & 0.1 & 15.60 & 0.02 &  9.0 &   87 & 6377776615769767936 \\
\hline
\end{tabular}
\smallskip
\\This table is available in its entirety in a machine-readable form in the online journal. A portion is shown here for guidance regarding its form and content.
\bigskip
\end{table*}

In this work we aim to verify the \emph{IUE}'s findings with a newer and larger sample of UV asteroid data from the {\it Galaxy Evolution Explorer} ({\it GALEX}), a NASA Small Explorer-class space telescope mission which from 2003--2012 conducted a UV imaging survey in a far-UV band (FUV, 130--190 nm) and a near-UV band (NUV, 180--280 nm). Approximately 2/3 of the sky was covered, with avoidance of bright stars and low galactic latitudes. \cite{mar05} discuss the extragalactic science program, while Morissey et al. (2005, 2007) discuss the on-orbit performance, survey calibration and data products. {\it GALEX} has a 50 cm$^2$ effective area, 1.25 degree diameter circular field of view, and full-width-half-maximum resolution of 4.5$''$ in the NUV. Programs within the {\it GALEX} mission included an all-sky survey (AIS, with $\sim$100 s exposures) and a medium-depth survey (MIS, with $\sim$1500 s exposures), and also a spectroscopic (grism) survey. Figure 1 shows the photometric response functions of the two \emph{GALEX} bandpasses multiplied by the solar spectrum, with comparison to the $ugriz$ visible bandpasses. Detection of asteroids in the FUV is extremely unlikely (nonetheless, as described below we searched for both NUV and FUV asteroid detections).

\begin{figure}[t]
\centering
\includegraphics[scale=0.88]{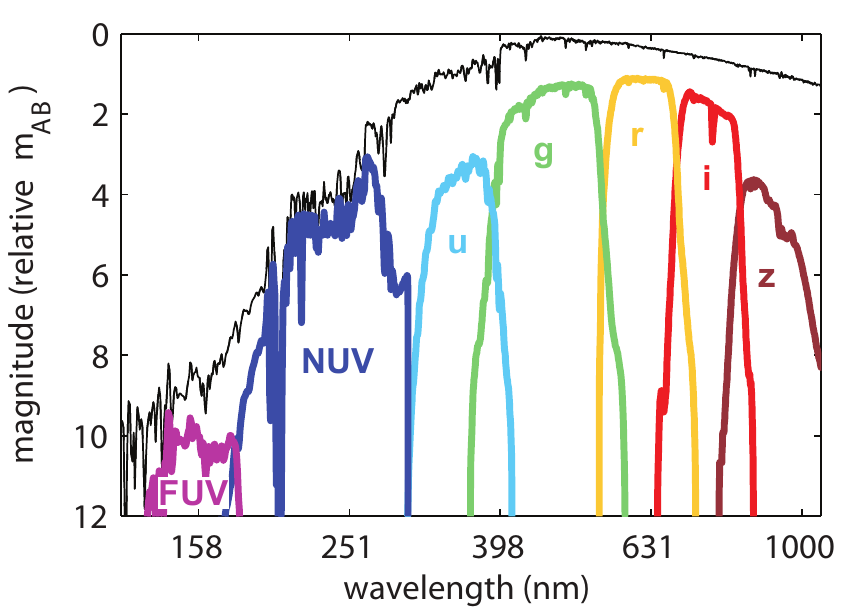}
\caption{{\it GALEX} UV and SDSS visible filter response curves (colored lines) multiplied by the spectrum of a G2 V type star (black line). The spectrum is from the library of \cite{pic98}. The vertical scale is in AB magnitude units per unit wavelength, offset by an arbitrary constant. Note wavelength is plotted on a log scale.}
\end{figure}

Our approach in analyzing \emph{GALEX} asteroid observations differs somewhat from Roettger and Buratti's treatment of the \emph{IUE} data. First, instead of referencing taxonomic class labels ({\it e.g.}, `C type',`S type',`M type', etc.) assigned to individual asteroids by previous authors, we define classes using a color index derived from a clustering analysis performed on a compilation of seven visible-color surveys \citep{was15}. The brightest asteroids typically were targeted in one or more spectroscopic surveys---{\it e.g.}, the Eight Color Asteroid Survey (ECAS, \citealp{zel85}) or the Small Main-Belt Asteroid Spectroscopic Surveys (SMASS; \citealp{xu95}, \citealp{busb02}). Dimmer objects however often only have color information from the Sloan Digital Sky Survey (SDSS; \citealp{yor00}, \citealp{ive01}, \citealp{par08}). The color index of \cite{was15} puts asteroids of all sizes on a single, quantitative color scale (a proxy for spectral slope), the endmembers of which we identify with the C-type and S-type complexes. We use the terms `C types' and `S types' purely for compatibility with the literature, noting that our color index combined with the classification thresholds we apply to it represent original definitions of these two groups.

Use of a one-dimensional color metric sacrifices sensitivity to secondary taxonomic groups such as M types, D types and V types. While our purpose is mainly to examine the \emph{first-order} taxonomic variation, in certain cases we divide our color-defined `C types' into two classes (C$_\text{high}$ and C$_\text{low}$) on the basis of visible and near-infrared albedo (as detailed in Section 3.2). At the end of this work we also examine well-established C-complex subgroups from both the Tholen and Bus/Binzel taxonomies (Section 6), showing how these known subgroups ({\it e.g.}, M types and G types) are distributed in this work's color-albedo space and highlighting these subgroups' unique UV properties.

A second distinction between this work's approach and \cite{roe94} is that, rather than comparing the \emph{geometric albedo} in the UV with that of the visible band, we focus on the difference in \emph{apparent magnitudes} between the UV and visible. One motivation for doing this is we need not make any assumptions about the phase function of asteroids in the UV. The challenge however is that we must accurately estimate the visible flux at the time of the UV observations. As discussed in Section 3, we adopt (and compare) two distinct methods for predicting the visual magnitude. The first method simply adopts the widely-used MPC\footnote{IAU Minor Planet Center, \href{http://minorplanetcenter.net}{\color{blue}{http://minorplanetcenter.net}}} predicted magnitudes; the second method applies color-dependent phase-function and bond-albedo estimates adapted from the \cite{was15} study of lightcurves from the Palomar Transient Factory survey\footnote{\href{http://ptf.caltech.edu}{\color{blue}{http://ptf.caltech.edu}}} (PTF; \citealp{law09}; \citealp{rau09}).

\begin{figure*}[t]
\centering
\includegraphics[scale=0.75]{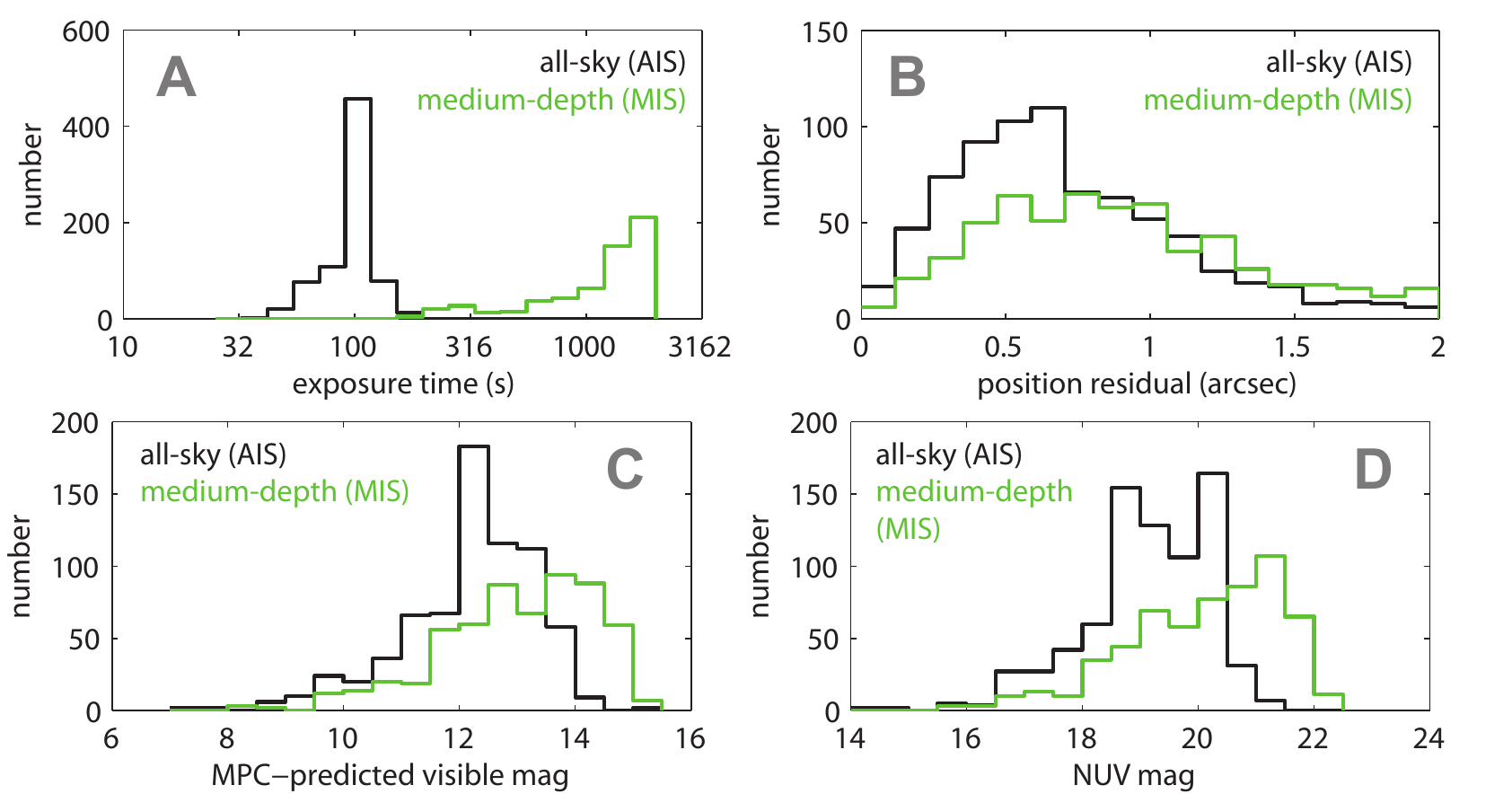}
\caption{Characteristics of positive asteroid detections from the two \emph{GALEX} surveys (distinguished by exposure time) shown separately in black and green.}
\end{figure*}

\section{GALEX asteroid observations}

Extracting detections of known asteroids from a survey involves a three-dimensional (R.A., Dec., time) cross-matching of ephemerides against the survey's time-stamped image boundaries ({\it e.g.}, \citealp{ofe12}). We modified software originally used to search for asteroids in PTF (\citealp{was13}, \citeyear{was15}) to instead search for asteroids in \emph{GALEX}.

We first retrieved the metadata of all {\it GALEX} images, available from the Space Telescope Science Institute via command-line queries with the CasJobs tool \citep{li08}. We then indexed all image centers with respect to (R.A, Dec.) into uniformly-spaced sky cells of 3-degree radius. For all $\sim$380,000 numbered asteroids, we queried JPL's online service HORIZONS \citep{gio96} to generate a 1-day-spaced ephemeris spanning 2003--2012. Using an object-specific search radius equal to 3 degrees (cell radius) plus 0.75 degrees (FOV radius) plus the object's maximum 1-day motion ($\sim$10 arcminutes for most main-belt objects), we matched the ephemeris points against the sky cells. For each matched cell, we filtered out all images in that cell not within the epoch range of the matched ephemeris points, then for each surviving image we re-queried HORIZONS for the precise location at each observed epoch. We next performed a 1.25-degree-radial match of these precise positions against the relevant {\it GALEX} image centers.

We found $\sim$850,000 predicted detections of numbered asteroids (with no limit on apparent magnitude) in {\it GALEX} using this method. For each predicted detection, using CasJobs we queried the {\it GALEX} single-visit source list (as opposed to the co-added source list). Multiple matches near the same point occurring more than 6 hours apart were excluded, as were all matches further than 2$''$ from the predicted location. Additionally, to ensure the inclusion of greater than (approximately) 5$\sigma$ detections, we discarded all matches with NUV $>$ 21 mag in the shorter exposures (AIS program), and discarded all matches with NUV $>$ 22.7 mag in the longer exposures (MIS program), following the limiting magnitudes quoted by \cite{mor07}.

Following the above procedure and criteria, we extracted a total of 1,342 positive NUV detections of 405 unique asteroids which were detected by {\it GALEX} at least twice (and no FUV detections, as expected). These detections are listed in Table 1; several histograms detailing these detections appear in Figure 2.

All of the \emph{GALEX}-observed asteroids are in the main-belt; the sample includes no near-Earth or outer-solar system objects.

\section{Modeling visible magnitudes}

In this section we consider two distinct methods of estimating the visible magnitudes corresponding to all {\it GALEX} NUV detections; this in turn provides the distribution of the asteroids' $\text{NUV} - V$ color. The general model for an asteroid's apparent visual magnitude $V$ (log flux) is

\begin{equation}
V=H+\delta+5\log_{10}(r\Delta)-2.5\log_{10}[\phi(\alpha)],
\end{equation}

\noindent where $H$ is the absolute magnitude (a constant), $\delta$ is a periodic variability term due to rotation ({\it e.g.}, if the object is spinning and has some asymmetry in shape or albedo), $r$ and $\Delta$ are the heliocentric and geocentric distances (in AU), and $\phi =\phi(\alpha)$ is the \emph{phase function}, which varies with the solar phase angle $\alpha$ (the Sun-asteroid-Earth angle). When $\alpha=0$ ({\it i.e.}, at opposition), $\phi=1$ by definition, while in general $0<\phi<1$ for $\alpha>0$ (with $\phi$ decreasing as $\alpha$ increases).

All asteroids for which we have extracted \emph{GALEX} observations have known orbits, meaning $r$, $\Delta$, and $\alpha$ are accurately and precisely known at all observed epochs. Our two methods for estimating $V$ differ in their assumptions regarding (and observational data used to constrain) $H$ and $\phi$. In both cases we do not attempt to model the rotational term $\delta$, but rather incorporate $\delta$ into the uncertainty of $V$ using lightcurve amplitude estimates from the literature. In particular, 388 of the 405 \emph{GALEX}-observed asteroids have an amplitude lower-limit estimate available in the Lightcurve Database (\citealp{war09}, \citealp{har12}).

In the following sections we refer to two different albedo quantities. The visible-band \emph{geometric} albedo $p_V$ relates to the visible-band \emph{bond} albedo $A_\text{bond}$ and the phase function $\phi$ (of Equation [1]) according to

\begin{equation}
p_V\equiv\frac{A_\text{bond}}{2}\left(\int_0^\pi \phi(\alpha)\sin(\alpha)\;d\alpha\right)^{-1}\equiv\frac{A_\text{bond}}{q},
\end{equation}

\noindent The above equation also defines the phase integral $q$. The bond albedo $A_\text{bond}$ is defined as the total visible light energy reflected or scattered by the asteroid (in all directions) divided by the total visible light energy incident upon the asteroid (from the Sun). Assuming the asteroid has a circular cross-section of diameter $D$, this can be expressed as

\begin{equation}
A_\text{bond}\equiv\frac{\int_0^\pi f(\alpha)\sin(\alpha)d\alpha}{(f_\text{Sun}/4\pi \text{AU}^2)\times\pi(D/2)^2},
\end{equation}

\noindent where $f(\alpha)=10^{-V(\alpha)/2.5}$ is the asteroid's flux as a function of phase angle, with $V(\alpha)=H-2.5\log_{10}\phi(\alpha)$ being Equation (1) evaluated at $\delta=0$ and $r=\Delta=1$ AU (similarly, $f_\text{Sun}=10^{-V_\text{Sun}/2.5}$).

\begin{figure*}[t]
\centering
\includegraphics[scale=0.69]{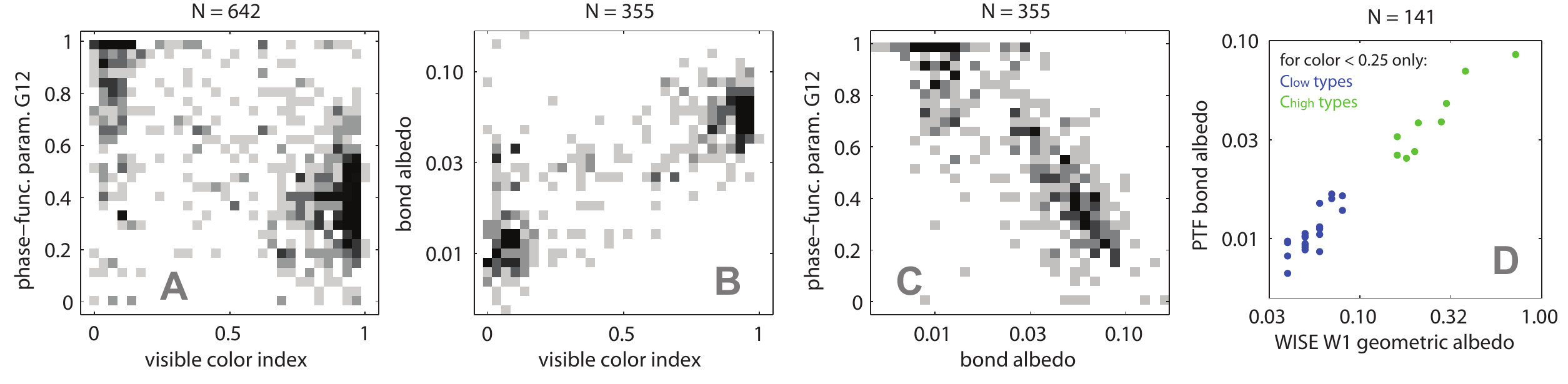}
\caption{We compute $V_\text{PTF}$ model magnitudes by first assigning fixed $A_\text{bond}$ and $G_{12}$ values to each \emph{GALEX}-observed asteroid depending on its color class; we then use $D$ to compute $H$, and finally use the assumed $G_{12}$ value to predict $V$. The fixed values of $A_\text{bond}$ and $G_{12}$ are medians from the color-albedo-$G_{12}$ data in \cite{was15}, 2D histograms of which are shown here. Above each plot is the sample size ($N=\ldots$). A total of 642 asteroids have color data \emph{and} $G_{12}$ values in the PTF data; 355 of these also have diameters available (required to compute $A_\text{bond}$). Panel D shows that \emph{WISE} $W1$ geometric albedos correlate with the PTF bond albedo; we thus use the \emph{WISE} $p_{W1}$ data to assign C types either a low ($A_\text{bond}\approx0.01$) or high ($A_\text{bond}\approx0.04$) bond albedo.}
\end{figure*}

\subsection{($H,G$) from MPC data}

The first method for estimating $V$ adopts the Minor Planet Center's computed absolute magnitudes ($H_\text{MPC}$), which are regularly updated by the MPC's automated processes and utilize the Lumme-Bowell $G$-parameter model for $\phi$ \citep{bow89}. This same ($H,G$) model then predicts the apparent magnitude $V_\text{MPC}$ as a function of solar phase angle.

The $H_\text{MPC}$ values are fit to photometry provided by a variety of surveys/individuals, many of whom may use slightly different absolute calibration standards or filters with slightly different specifications. A small fraction of asteroids have fitted $G$ values; \cite{hary88} present mean $G$ values for several major taxonomic classes, with $G=0.15$ being an average between the C types ($G\approx 0.08$) and the S types ($G\approx 0.23$). For the majority of asteroids the MPC uses an assumed $G=0.15$ with this model. \cite{was15} compares the $H_\text{MPC}$ values with $H$ magnitudes derived from a model that includes rotation and the more modern ($H$,$G_{12}$) phase function of \cite{mui10}. Among bright asteroids the relative difference is typically between 0.3\% to 3\%, corresponding to (on average) an $\sim$0.07 mag discrepancy.

Though $H_\text{MPC}$ values are available for all 405 \emph{GALEX}-observed asteroids, we only consider the subset of 315 asteroids having visible-band color indices of either less than 0.25 (`C types') or greater than 0.75 (`S types'). Of these, 41 asteroids have $G_\text{MPC}\ne0.15$.

\subsection{($D$,$A_\text{bond}$,$G_{12}$) from PTF, infrared, and color data}

Our second means of estimating visual magnitudes applies only to asteroids having both a color index \emph{and} a diameter estimate constrained from thermal fluxes in an infrared survey\footnote{Similar to the color data, the diameter data set we use is a compilation of products from several surveys and described in the appendix of \cite{was15}. The source IR surveys are \emph{WISE} (\citealp{wri10}, \citealp{mas11}, \citealp{mas14}), \emph{IRAS} (\citealp{mat86}, \citealp{ted02a}), \emph{MSX} \citep{ted02b}, and AKARI \citep{usu11}.}. In this approach we use the $G_{12}$-parameter model for $\phi$ \citep{mui10}, and we replace $H$ with its equivalent expression\footnote{Equation (4) follows \emph{directly} from combining Equations (1)--(3). The constant 1329 km depends on somewhat arbitrarily-defined quantities such as the Sun's visual magnitude and the ratio of an AU to a kilometer.} in terms of the diameter $D$, bond albedo $A_\text{bond}$, and phase integral $q$:

\begin{equation}
H=-5\log_{10}\left(\frac{D\sqrt{A_\text{bond}/q}}{\text{1329 km}}\right),
\end{equation}

\noindent where the phase integral $q$ is a linear function of $G_{12}$:

\begin{equation}
q(G_{12}) =\left\{
\begin{split}
&0.2707-0.236 G_{12}\;\;\;\text{if}\;G_{12}<0.2;\\
&0.2344-0.054 G_{12}\;\;\;\text{otherwise}.
\end{split}
\right.
\end{equation}

We again define `C types' as all asteroids with color indices less than 0.25 and `S types' as all with color indices greater than 0.75. For S types we then consider diameters derived from any of four infrared surveys (see Footnote 3), while for C types we specifically require that the asteroid have been observed in the \emph{WISE} 4-band cryogenic survey (\citealp{wri10}, \citealp{mas14} and references therein). Both the \emph{WISE} $W1$-band geometric albedo $p_{W1}$ and the PTF-derived bond albedo\footnote{The visible bond albedo $A_\text{bond}$ uses the same \emph{WISE} diameter used by \cite{mas14} in computing the $W1$-band geometric albedo $p_{W1}$.} $A_\text{bond}$ show evidence of bimodality among objects with color indices less than 0.25 (Figure 3 panel D). Thus, we divide the C types into low-bond-albedo (C$_\text{low}$) and high-bond-albedo (C$_\text{high}$) subgroups based on their $p_{W1}$ as reported by \cite{mas14}. In Section 6 we show that the C$_\text{high}$ types most closely correspond to what other authors have called M types.

\cite{was15} computed $A_\text{bond}$ and $G_{12}$ values for $\sim$1,600 asteroid lightcurves in the PTF survey. Using that work's data (Figure 3) we compute median $A_\text{bond}$ and $G_{12}$ values (and associated scatter) for the S, C$_\text{low}$ and C$_\text{high}$ taxonomic groups. Table 2 summarizes the definitions and assumed $A_\text{bond}$ and $G_{12}$ values of these groups. There are 245 \emph{GALEX}-observed  asteroids (out of the 405 in Table 1) which have color \emph{and} diameter data available, allowing them to be modeled by this method. To each \emph{GALEX}-observed asteroid we assign the appropriate $A_\text{bond}$ and $G_{12}$ value based on its class membership, then use its diameter to compute a model absolute magnitude ($H_\text{PTF}$) using Equation (4). Together with the assumed $G_{12}$ value, this $H_\text{PTF}$ then predicts the apparent magnitude $V_\text{PTF}$ at each \emph{GALEX}-observed solar phase angle.

\begin{table}
\caption{$A_\text{bond}$ and $G_{12}$ (based on PTF data) of color-defined taxonomic groups.}
\begin{tabular}{cccccccccc}
\hline\\[-2ex]
 class    & color  & \emph{WISE} & $A_\text{bond}$     &  $A_\text{bond}$ & $G_{12}$ & $G_{12}$\\[-0.3ex]
 name     & index  & $p_{W1}$    & median  & scatter$^*$ &    median & scatter$^*$\\
\hline\\[-2ex]
 S              & $>0.75$ &   N.A.             & 0.056 & 0.016 & 0.36 & 0.16 \\
 C$_\text{high}$& $<0.25$ & $>0.125$ & 0.038 & 0.022 & 0.42 & 0.20  \\
 C$_\text{low}$& $<0.25$ & $<0.125$ &  0.010 &  0.003 & 0.84 & 0.16  \\
\hline
\end{tabular}
\smallskip
\\$^*$Scatter is here defined as $0.5\times(84^\text{th}$ percentile $-16^\text{th}$ percentile)
\bigskip
\end{table}

\begin{figure*}[t]
\centering
\includegraphics[scale=0.69]{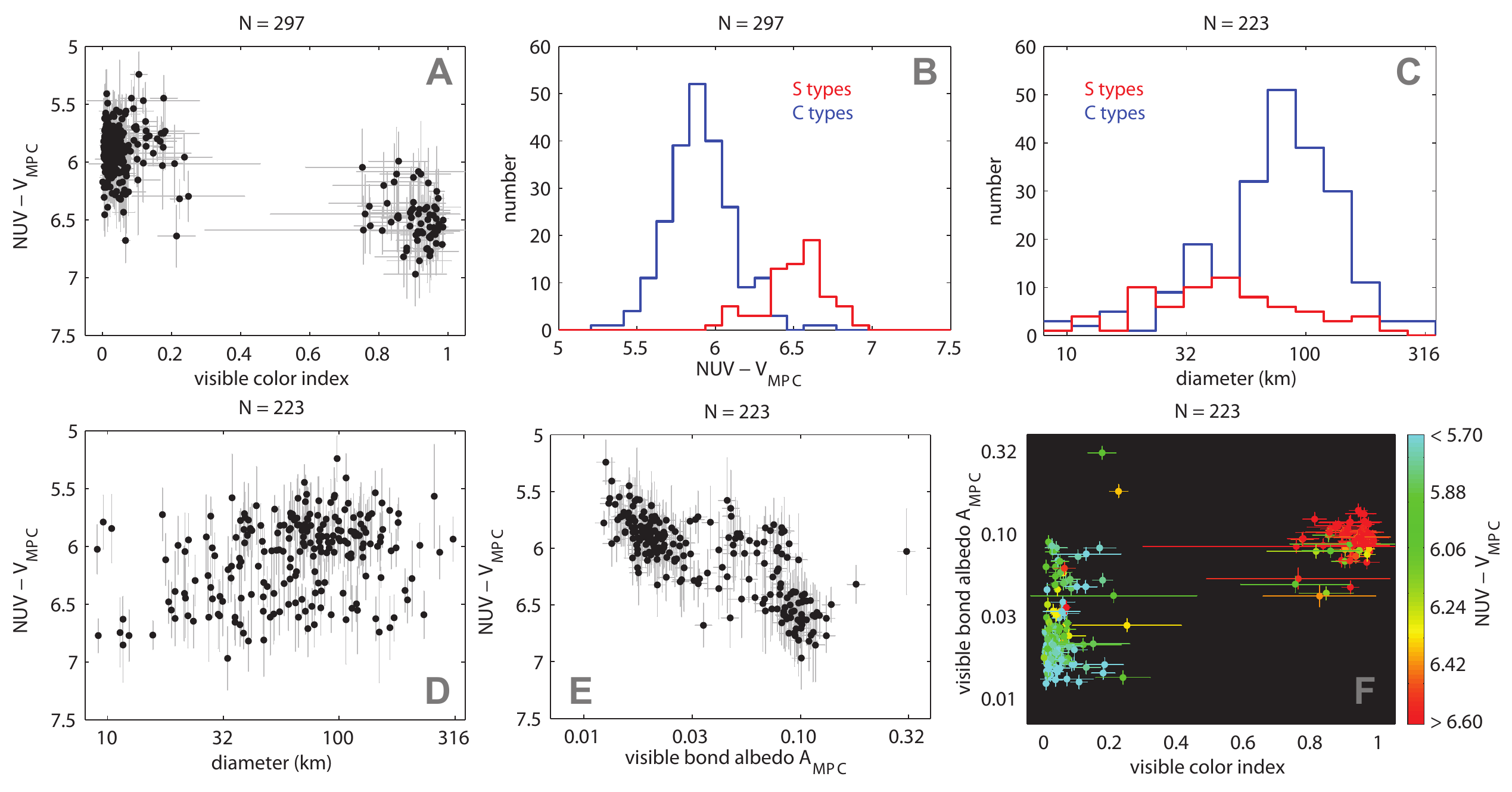}
\caption{Distribution of the $\text{NUV}-V$ color for \emph{GALEX}-observed asteroids using the $H,G$ model with MPC data to predict $V$. Plots B and C define C types and S types as objects with color indices of $<$0.25 and $>$0.75, respectively. Plots C--F include only the subset with a diameter estimate available; this subset is precisely the same sample considered in Figure 6.}
\vspace{10pt}
\end{figure*}

\begin{figure}
\centering
\includegraphics[scale=0.7]{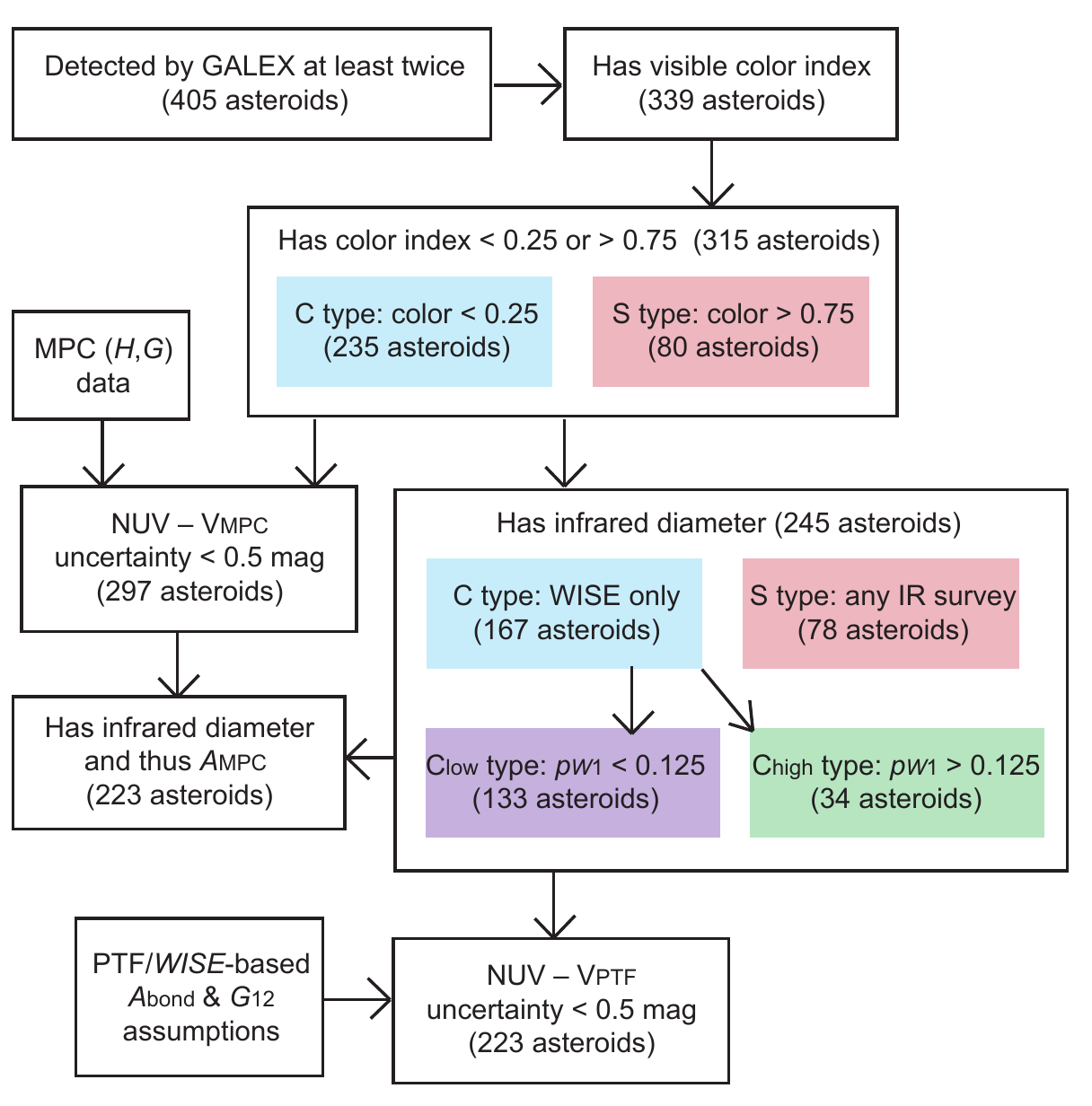}
\caption{Flowchart visualizing the steps in the \emph{GALEX}-observed asteroid sample selection process. Each box is a subset of the box pointing to it.}
\vspace{10pt}
\end{figure}

\begin{figure*}[t]
\centering
\includegraphics[scale=0.69]{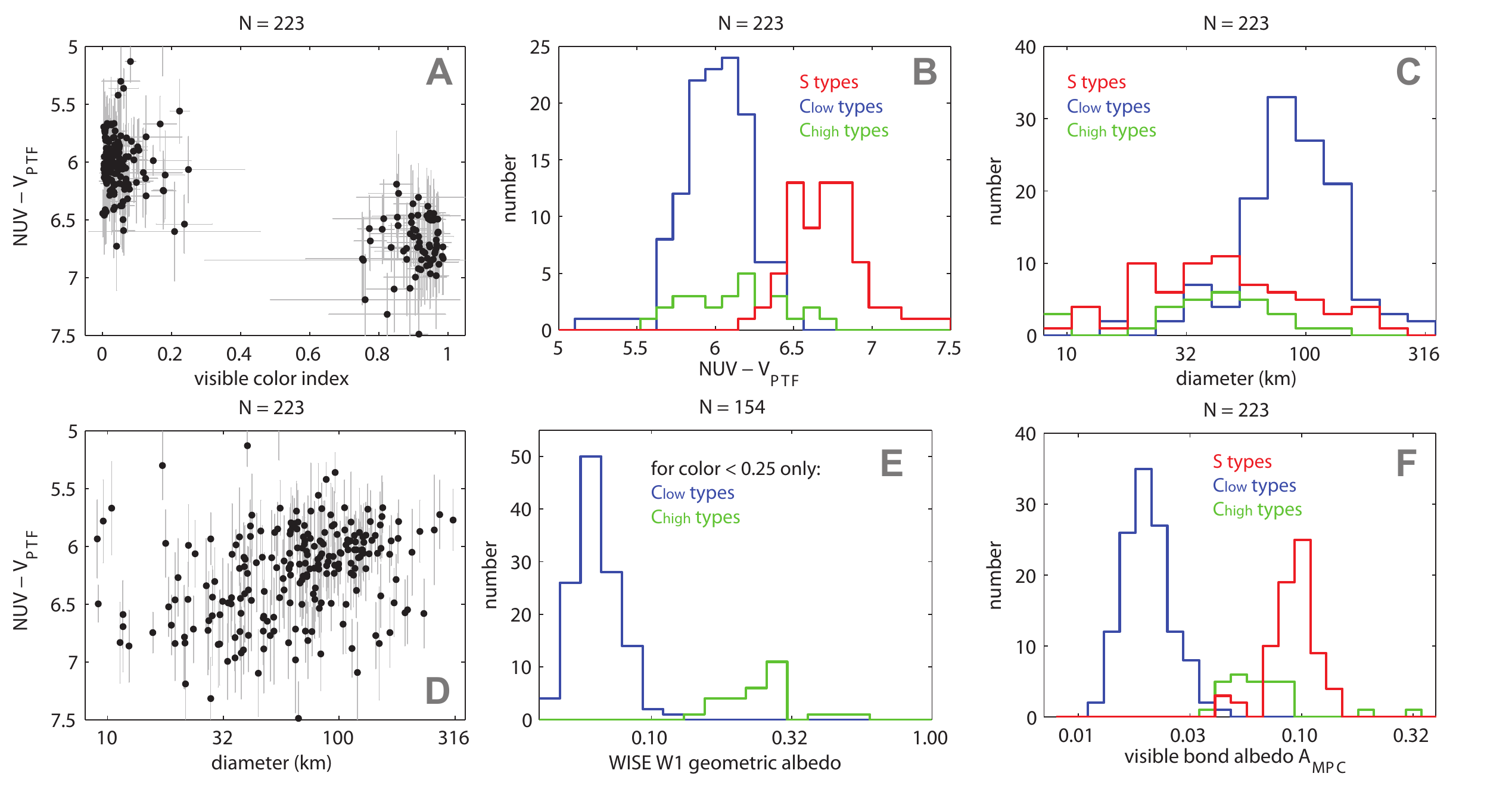}
\caption{Distribution of the $\text{NUV}-V$ color for \emph{GALEX}-observed asteroids using the $D$,$A_\text{bond}$,$G_{12}$ model with PTF data, infrared data, and color data to to predict $V$. See Table 2 for the definitions of the S, C$_\text{low}$, and C$_\text{high}$ groups in plots B, C, E, and F.}
\vspace{10pt}
\end{figure*}

\subsection{Rotational uncertainty in $V$}

Both the $V_\text{MPC}$ and $V_\text{PTF}$ model magnitudes discussed here lack an estimate of the rotational term ($\delta$ in Equation [1]). We account for this by incorporating a term for rotational modulation into the reported uncertainty of $V$. Of the 315 asteroids with $V_\text{MPC}$ values, 302 have an amplitude lower limit listed in the Lightcurve Database (\citealp{war09}, \citealp{har12}) , while for the 245 asteroids with $V_\text{PTF}$ predictions there are 239 with reported amplitudes. As shown for instance by \cite{was15}, asteroids in the relevant size range typically have amplitudes less than $\sim$0.4 mag. For the few objects in our sample lacking an amplitude limit, we assume a value of 0.2 mag.

Assuming an asteroid's rotational phase $\varphi$ to be random at the time of a \emph{GALEX} detection ({\it i.e.}, with a probability distribution of the form $P(\varphi)\propto$ constant), then the probability distribution of a basic sinusoidal $\delta$ ({\it i.e.}, one of the form $\delta=\delta_0 \sin\varphi$) can be shown to have the form

\begin{equation}
P(\delta)\propto \frac{1}{\sqrt{\delta_0^2-\delta^2}},
\end{equation}

\noindent where $\delta_0$ is the amplitude. We use Equation (6) as a probability density function to generate, for each modeled $V$, a set of $10^4$ simulated $\delta$ values. These simulated $\delta$ are added to an equal number of model $V$ magnitudes computed by random (Gaussian distribution) sampling of the component terms: in the case of $V_\text{MPC}$ we just assume a fixed $H_\text{MPC}$ uncertainty of 0.1 mag, whereas for the $V_\text{PTF}$ values we randomly sample all three of $A_\text{bond}$, $G_{12}$, and $D$, using the scatter values in Table 2 for the first two and the literature-reported diameter uncertainty for $D$. The 16$^\text{th}$ to 84$^\text{th}$ percentile spread in the distribution of combined $\delta+V$ values then becomes the quoted uncertainty for $V$.

\section{$\text{NUV} - V$ color distribution}

Having computed the model $V$ magnitudes, we obtain the $\text{NUV} - V$ color for each \emph{GALEX} asteroid detection and the corresponding uncertainty. The latter contains an additional rotational uncertainty component (now associated with the NUV observation), again determined by repeated sampling of Equation (4) as described above. Since all the asteroids we consider have more than one \emph{GALEX} NUV detection, we compute the variance-weighted average $\text{NUV} - V$ color for each asteroid (plotted in Figures 4 and 6); the uncertainty in this average is the inverse quadrature sum of the individual uncertainties.

In Figures 4 and 6 (and the accompanying analysis) we have omitted all asteroids with $\text{NUV} - V$ uncertainties of greater than 0.5 mag. As a result, the sample size of asteroids with $\text{NUV} - V_\text{MPC}$ estimates is 297 (out of the 315 quoted in Section 3.1), while the sample with $\text{NUV} - V_\text{PTF}$ estimates is 223 (out of the 245 quoted in Section 3.2). Figure 5 graphically summarizes the sample selection criteria in a flowchart. In Figures 4, 6 and 7, the errorbars on the color indices were computed by a bootstrapping process described in the appendix of \cite{was15}. 

\begin{table}
\caption{$\text{NUV}-V$ color (mag. units) of \emph{GALEX}-observed asteroids and sample sizes}
\begin{tabular}{cccccccccc}
\hline\\[-2ex]
 class    & \multicolumn{2}{c}{$\text{NUV}-V_\text{MPC}$}   & \multicolumn{2}{c}{$\text{NUV}-V_\text{PTF}$} & \multirow{2}{*}{$N_\text{MPC}$} & \multirow{2}{*}{$N_\text{PTF}$} \\[-0.3ex]
 name     & median  &   scatter$^*$   & median  & scatter$^*$ &  & &\\
\hline\\[-2ex]
 S              & 6.52 & 0.25 & 6.71 & 0.21 & 72 & 69 \\
 C              & 5.90 & 0.19 & 6.03 & 0.22 & 225 & 154 \\
 C$_\text{high}$&  --  & --   & 6.14 & 0.33 & -- & 29  \\
 C$_\text{low}$ &  --  & --   & 6.02 & 0.19 & -- & 125  \\
\hline
\end{tabular}
\smallskip
\\$^*$Scatter is here defined as $0.5\times(84^\text{th}$ percentile $-16^\text{th}$ percentile)
\end{table}

Both the $V_\text{MPC}$ and $V_\text{PTF}$ model magnitudes produce a bimodal $\text{NUV} - V$ color distribution, with the S types having the redder $\text{NUV} - V$ color (panels A and B of both Figures 4 and 6). Median and scatter of $\text{NUV} - V$ for the various classes appear in Table 3. To formally ascertain the inequality of the two distributions, we use the two-sided Kolmogorov-Smirnov (KS) test \citep{mas51}, which compares two empirical distributions via a bootstrap method. In particular this test computes a statistic quantifying the extent to which the cumulative distribution function differs in the two distributions being compared. For the $V_\text{MPC}$ model we find the C-type $\text{NUV}-V$ color distribution differs from that of the S-type distribution at an $11.6\sigma$ significance level (Figure 4 panel B). For the $V_\text{PTF}$ model (Figure 6 panel B) we find the C types (C$_\text{low}$ and C$_\text{high}$ combined) differ from the S types at an $8.1\sigma$ level, while the C$_\text{low}$ and C$_\text{high}$ types only differ at a $1.9\sigma$ level (this difference is thus not statistically significant). 

An important characteristic of our sample is that the C types outnumber the S types by a ratio of 3:1 in the $V_\text{MPC}$ sample and a ratio of 2:1 in the $V_\text{PTF}$ sample (cf. panel C of Figures 4 and 6). This ratio is a combination of (1) the inherent difference in the population sizes of the two types (above a given diameter cut-off), a detection bias due to S types dominating the inner main-belt and thus typically having brighter apparent magnitudes for a given size and albedo, and (3) the difference in the S and C types' NUV albedo (discussed in Section 5).

In panels C--F of Figure 4 the sample size decreases from $N=297$ down to $N=223$ asteroids as we consider only those objects in the $V_\text{MPC}$ sample that also have available diameters (this is equivalently the $V_\text{PTF}$ sample considered in Figure 6). We compute the MPC-based visible bond albedo $A_\text{MPC}$ using Equation (2) together with the asteroid's $H_\text{MPC}$ and $G_\text{MPC}$ values. In particular, there are 38 asteroids (out of the 223 with diameters) with a measured $G_\text{MPC}\ne0.15$; for the remainder we assume $G_\text{MPC}=0.15$ for consistency with the manner in which the $V_\text{MPC}$ are computed. Analogous to Equation (5), the phase integral for the $G$-model (required for computation of $A_\text{MPC}$ via Equation [4]) is

\begin{figure*}[t]
\centering
\includegraphics[scale=0.69]{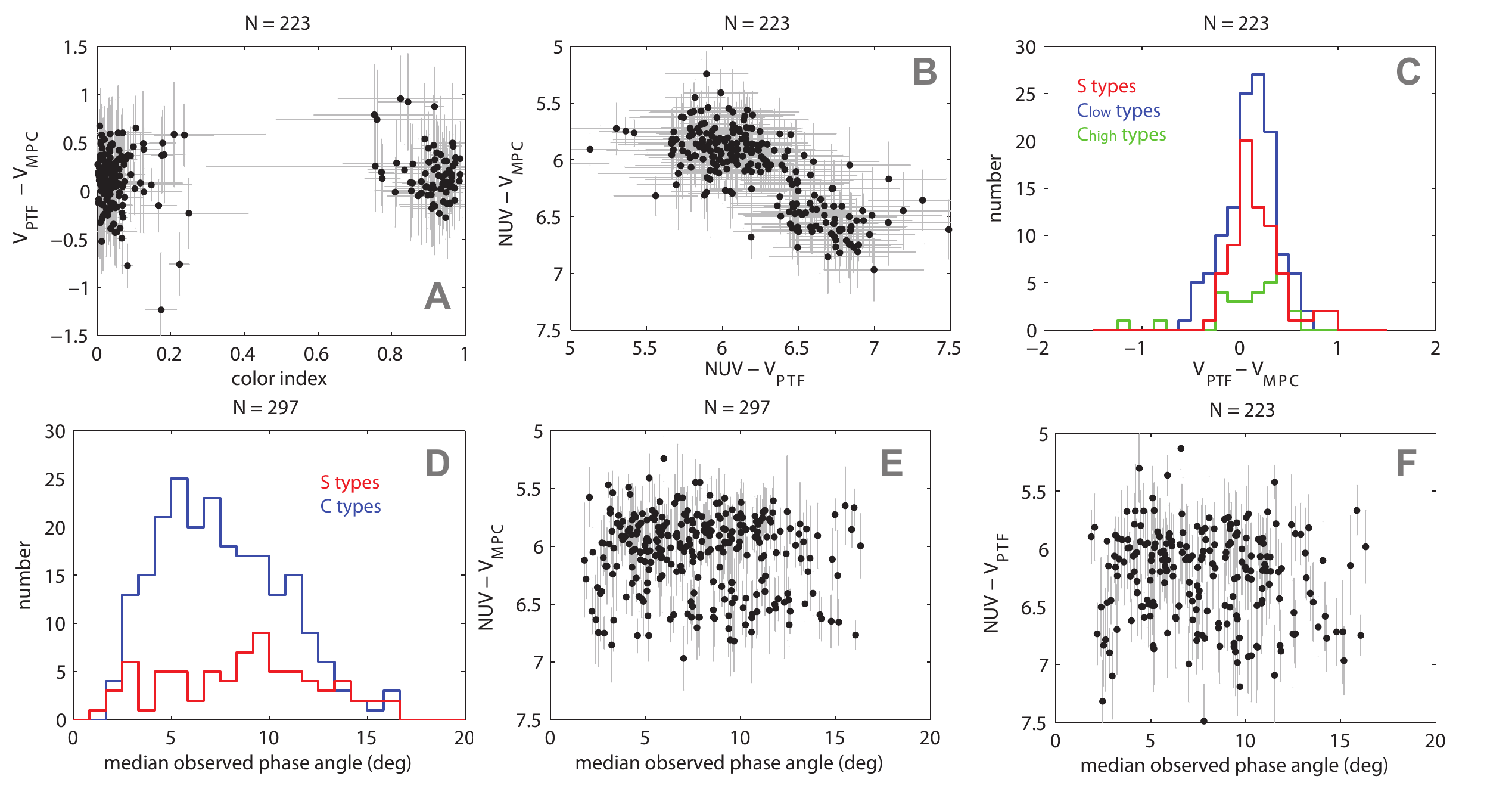}
\caption{Various checks for systematic differences in the predicted $V$ magnitudes output by the two different photometric models. \emph{Bottom row}: Investigation of phase-angle-dependence on the $\text{NUV}-V$ color.}
\vspace{10pt}
\end{figure*}

\begin{equation}
q(G) = 0.290 + 0.684 G,
\end{equation}

\noindent as given by \cite{bow89}. With the $G_\text{MPC}=0.15$ assumption for the majority of the asteroids in our sample, the $A_\text{MPC}$ values are not expected to be as accurate as the $A_\text{bond}$ values computed for instance by \cite{was15}, wherein distinct $q$ values were fitted to each object on the basis of a lightcurve. Nonetheless, it is instructive to compute $A_\text{MPC}$, {\it e.g.},  to check for consistency with the class-median $A_\text{bond}$ values, and to exploit as a second taxonomic metric in addition to visible color.

Figure 4 panel E shows that $\text{NUV} - V$ correlates with $A_\text{MPC}$ ($\rho_\text{Spearman}=0.698$, $>$$10\sigma$ significance), similar to how $\text{NUV} - V$ correlates with the color index in panel A ($\rho_\text{Spearman}=0.491$, $>$$10\sigma$ significance). Unlike the color index however, the separation between the C$_\text{low}$ and C$_\text{high}$ subgroups is qualitatively evident in this plot. Figure 4 panel F combines all three parameters; note the axes are the same as Figure 3 panel B, with $A_\text{MPC}$ replacing $A_\text{bond}$ and the data consisting of \emph{GALEX}-observed asteroids rather than PTF-observed asteroids.

Figure 6 panel F confirms (independently of Figure 3 panel D) the validity of using \emph{WISE} $W1$-band geometric albedo as a proxy for visible bond albedo to separate C$_\text{low}$ from C$_\text{high}$---the two classes robustly differ in their $A_\text{MPC}$ distributions ($9.5\sigma$ KS-test significance). However, the class-median $A_\text{MPC}$ values of the C$_\text{low}$, C$_\text{high}$, and S types are 100\%, 67\%, and 63\% \emph{greater} than their class-median PTF-based $A_\text{bond}$ values in Table 2. This reflects the differing values of $H$ and $q$ produced by the $G$ and $G_{12}$ models, as well as the fact that we apply class-specific $G_{12}$ values, whereas $G_\text{MPC}=0.15$ is assumed for the majority of asteroids, regardless of their class.

Consideration of both the $V_\text{MPC}$ and $V_\text{PTF}$ model magnitudes provides two independent means of computing $\text{NUV} - V$; this helps rule out the effect of potential systematic errors unique to either one of the $V$ models, as well as possible biases in the distinct observational data sets upon which each $V$ is based. In Figure 7 panels A--C we examine the distribution of $V_\text{MPC}-V_\text{PTF}$ for all 223 asteroids having both $V$ estimates. The median of $V_\text{MPC}-V_\text{PTF}$ is 0.13 mag (scatter of 0.25 mag), indicating the MPC-based model consistently produces brighter $V$ estimates. For C types the median $V_\text{MPC}-V_\text{PTF}$ is 0.14 mag and for S types it is 0.12 mag; the two groups' $V_\text{MPC}-V_\text{PTF}$ distributions differ with less than $0.1\sigma$ significance in a KS-test.

In Section 1 we motivated our choice to examine the difference in \emph{apparent magnitude} between UV and visible (as opposed to the difference in \emph{albedo} in UV and visible) by noting that little is known of asteroid phase functions in the UV, rendering difficult the estimation of UV absolute magnitudes (and hence UV albedos). A potential issue with this approach which we have heretofore ignored is that, if the phase function \emph{does} differ significantly in the UV from the visible, then the $\text{NUV} - V$ color will vary with phase angle. Figure 7 panels D--F attempt to ascertain whether such a trend exists by considering the median phase angle at which each asteroid was detected by \emph{GALEX}.

C types are observed at a median median phase angle of 7.2 deg compared to the S types' median median phase angle of 9.0 deg. This is explained by the fact that C types on average have larger semi-major axes, which geometrically correspond to lower observed phase angles from Earth. Within the C-type group, median phase angle correlates with $\text{NUV} - V_\text{MPC}$ at $\rho_\text{Spearman}=-0.1$ (1.5$\sigma$ significance) and with $\text{NUV} - V_\text{PTF}$ at $\rho_\text{Spearman}=0.01$ (0.1$\sigma$ significance). Among S types, median phase angle correlates with $\text{NUV} - V_\text{MPC}$ at $\rho_\text{Spearman}=0.01$ (0.1$\sigma$ significance) and with $\text{NUV} - V_\text{PTF}$ at $\rho_\text{Spearman}=0.07$ (0.5$\sigma$ significance). We therefore cannot claim any phase angle dependence for $\text{NUV} - V$, regardless of the taxonomic group or $V$-model being considered.

Various works ({\it e.g.}, \citealp{san12} and references therein) discuss the phenomenon of asteroid phase reddening, {\it i.e.} an observed reddening of visible color with increasing phase angle. Very few survey-scale samples have been used to test for the presence of this effect. \cite{sza07} computed slightly different phase-angle dependences for the $g-r$ and $r-i$ colors of Trojans in SDSS, though these relations were not separately computed for the Trojans' two taxonomic groups. \cite{was15} did not detect any statistically significant difference between $G_{12}$ fits to $r$-band PTF lightcurves and $g$-band PTF lightcurves (among asteroids that had data in both bands). The extent to which a phase-function dependence on wavelength exists between the UV and visible remains unclear. Future UV surveys such as \emph{ULTRASAT} \citep{sag14} offer the most promising means of testing this hypothesis, especially because (unlike \emph{GALEX}) they will obtain sufficient numbers of observations to adequately sample UV lightcurves, thereby providing the best possible data set for fitting UV phase functions.

\section{Albedo vs. wavelength}

If we assume that the phase function does not differ significantly between the UV and visible (or take Figure 7 panels E and F as justification of this statement), then we can compare the relative \emph{bond albedo} versus wavelength for the different taxonomic groups using measured colors, filter response functions and the solar spectrum. If the phase function \emph{does} in fact vary significantly with wavelength, then this approach only provides the relative \emph{geometric albedo} versus wavelength (see Equations [2] and [3]).

Assume photometry from two filters (1 and 2) produce the color measurement $m_1-m_2$. This color relates to the solar flux distribution $S(\lambda)$, the albedos in each band ($A_1$ and $A_2$) and the filter responses $F_1(\lambda)$ and $F_2(\lambda$) according to

\begin{equation}
10^{(m_1-m_2)/2.5}=\frac{\int F_1(\lambda)\lambda^{-2} d\lambda\int S(\lambda)F_2(\lambda)A_2 d\lambda}{\int S(\lambda)F_1(\lambda)A_1 d\lambda \int F_2(\lambda)\lambda^{-2} d\lambda}
\end{equation}

\noindent which we adapted from a similar equation in \cite{pic98}.

Using NUV as band 1 and $V$ as band 2 in Equation 6, we use the colors in Table 3 (specifically, the $V_\text{PTF}$-based colors) to obtain the albedo ratio $A_\text{NUV}/A_V$, with uncertainties coming from the associated scatter in the colors. In Figure 7 we plot these albedo ratios for the C types and S types, incorporating an additional uncertainty component from the transformation from $r$ to $V$ (see \citealp{was15} for a discussion of this transformation in the context of asteroids). The end-computed values are $(A_\text{NUV}/A_r)_\text{C}=0.63_{-0.12}^{+0.14}$ and $(A_\text{NUV}/A_r)_\text{S}=0.33_{-0.06}^{+0.07}$. The relative albedo values in the SDSS bands included for comparison in Figure 8 are taken directly from a figure in \cite{ive01}; likewise the ECAS data are taken directly from a figure in \cite{zel85}.

Note that Figures 8 and 9 ignore the fact that the S types' and C types' \emph{absolute albedo} in $r$-band differs.  In other words, these plots could be converted into ones with absolute albedo on the vertical scale by multiplying the blue and red lines by their respective absolute $r$-band albedos, which would be similar to those listed in Table 2 for $V$ band. Assuming (from Table 2) that C types have $p_V=0.054\pm0.019$, then $(p_\text{NUV})_\text{C}=0.035\pm0.019$, and assuming S types have $p_V=0.264\pm0.031$, then $(A_\text{NUV})_\text{S}=0.088\pm0.028$. That is, S types have a higher mean NUV albedo than C types, similar to the visible.

Both C and S types show a continued trend of decreasing albedo at shorter wavelengths. Whereas for S types this behavior was already well-established in the 300--800 nm region, for C types the $u-g$ color had previously represented a significant deviation from the shallower slope observed from 400--800 nm. The C-type $\text{NUV}$ albedo in Figure 8 confirms the presence of a marked drop in albedo somewhere in the 200--400 nm range. Given the resolution of Figure 8 and the uncertainties in the data points, we cannot judge whether the C-type albedo levels off between NUV and $u$ band, or whether the slope between $u$ and $g$ bands persists into these shorter wavelengths. The \emph{IUE} spectra from \cite{roe94} indicate C-type albedo is constant at least in the range 240--300 nm (as does the Lutetia data described below), so that the former interpretation may be more accurate.

\begin{figure}[t]
\centering
\includegraphics[scale=0.95]{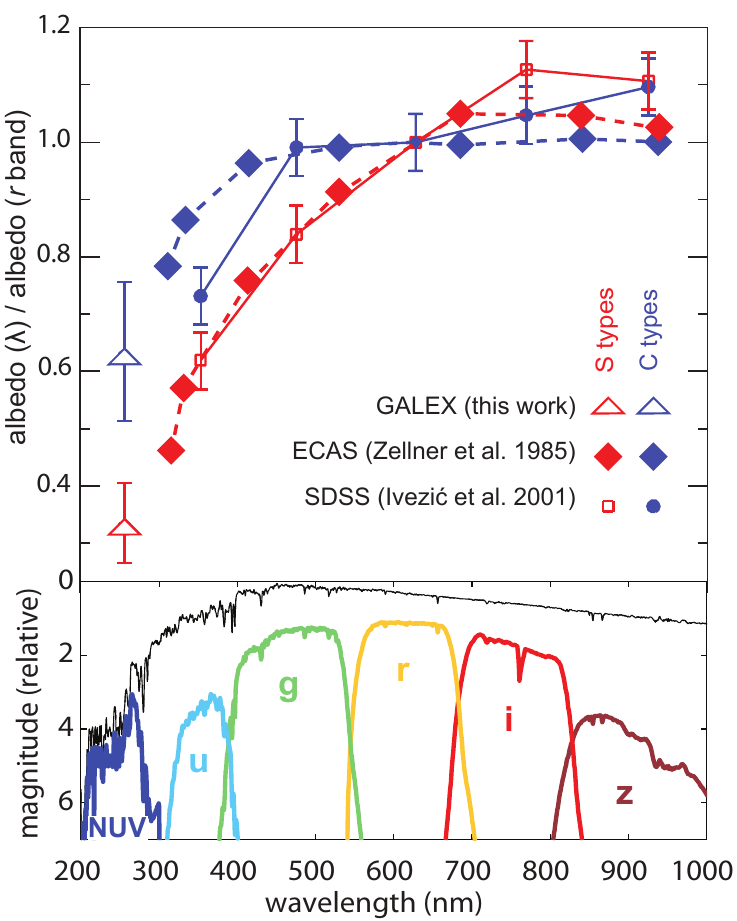}
\caption{\emph{Top}: Relative albedo (each type's $r$-band albedo normalized to unity) versus wavelength for C types and S types. The leftmost (NUV) points are computed from this work's data, the remaining albedos (SDSS bands) are taken directly from \cite{ive01}. \emph{Bottom}: Bandpass response functions (colored lines) multiplied by the solar spectrum (black line).}
\vspace{5pt}
\end{figure}

\begin{figure}[t]
\centering
\includegraphics[scale=0.95]{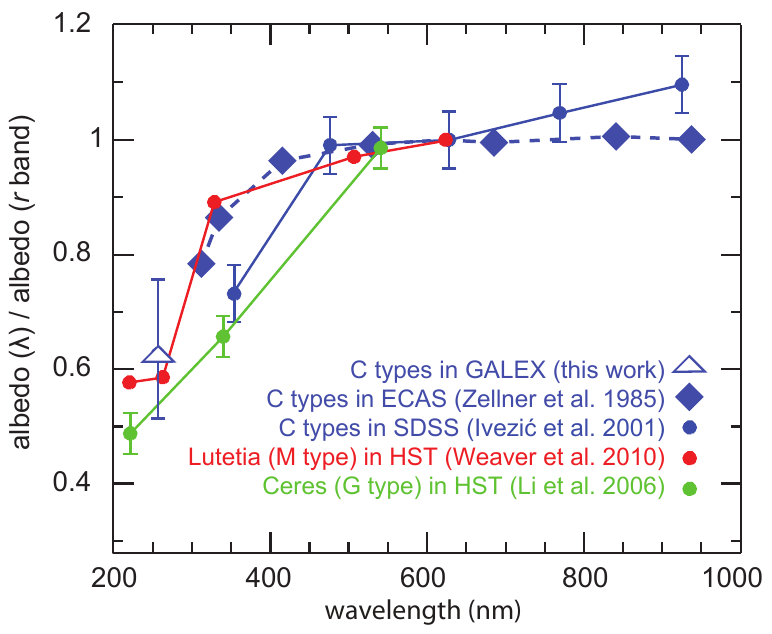}
\caption{\emph{Top}: \emph{GALEX}, SDSS, and ECAS C-type data from Figure 8 compared to \emph{HST}-derived albedos for 1 Ceres and 21 Lutetia. Note the red color here has a different meaning than it does in Figure 8. }
\vspace{5pt}
\end{figure}

\section{Comparison to \emph{HST} data}

\subsection{Lutetia}

\cite{wea10} obtained \emph{HST} photometry of asteroid 21 Lutetia in UV and visible bands. Lutetia has been classified by various authors as an M-type asteroid; in the context of this work its color index is 0.05 (making it a C type) and its $A_\text{MPC}=0.06$ suggest it to be a C$_\text{high}$ type in particular, though in this work's system we formally would require a \emph{WISE} $p_{W1}$ measurement to classify it as such. In the following section we show that M types (a group in the Tholen taxonomic system) and our C$_\text{high}$ types are largely the same population.

The \emph{HST} Lutetia photometry revealed a steep drop in albedo around $\sim$300 nm and nearly constant albedo in the 200--300 nm region at a factor $\sim$0.6 times the visible ($r$-band-equivalent) albedo. The \emph{HST} observations of Lutetia thus generally agree with the C-type albedo trend (Figure 9), the main difference being the location of the UV albedo drop-off (the bluest two ECAS bands also demonstrate this difference between M types and C types, {\it e.g.}, see Figure 2 of \citealp{bus02}). The \emph{Rosetta} spacecraft's flyby of Lutetia enabled FUV observations with the on-board Alice UV imaging spectrograph \citep{ste11}; the longest wavelengths of the FUV data ($\sim$190 nm) yield an albedo consistent with the constant value measured in the 200--300 nm range by \emph{HST}, namely $\sim$0.6 times that of the visible albedo.

\subsection{Ceres}

\emph{HST} photometry of asteroid 1 Ceres has also been obtained in the UV and visible (\citealp{par02}, \citealp{li06}). With a color index of 0.01, Ceres is also a C type in our classification scheme, though its $A_\text{MPC}=0.033$ makes its placement in our C$_\text{low}$ vs. C$_\text{high}$ groups ambiguous (see Figure 6 panel F). Like Lutetia, Ceres lacks a reported $p_{W1}$ so that we cannot formally classify it as either C$_\text{low}$ or C$_\text{high}$.

Ceres was observed by \emph{GALEX} and thus is included in our MPC-data-based analysis; our measured $\text{NUV}-V_\text{MPC}=6.45\pm0.19$ for Ceres make it a clear outlier from the C-type $\text{NUV}-V_\text{MPC}$ distribution (Figure 4 panel B). In the Tholen taxonomic system Ceres is classified as a G type; in the following section we show that other G types exhibit similarly high $\text{NUV}-V_\text{MPC}$ values but less anomalous $\text{NUV}-V_\text{PTF}$.

The \cite{par02} \emph{HST} data show that around $\sim$300 nm Ceres' albedo drops to as low as $\sim$0.3 times the visible-band albedo---compared to the factor of $\sim$0.6 seen for \emph{GALEX} C types and the Lutetia data---but that around $\sim$200 nm it appears to rise again to a more typical C-type UV albedo. \cite{roe94} did not observe this unusually deep absorption feature near 300 nm in their \emph{IUE} spectrum of Ceres; if real this feature could partially explain the anomalous $\text{NUV}-V_\text{MPC}$ we observe for G types in \emph{GALEX}. Figure 9 shows Ceres data in the three \emph{HST} bands observed by \cite{li06}, none of which sample the 300-nm region containing the putative absorption band, though these three bands do generally match the \emph{GALEX} C-type data.

\begin{figure*}
\centering
\includegraphics[scale=0.74]{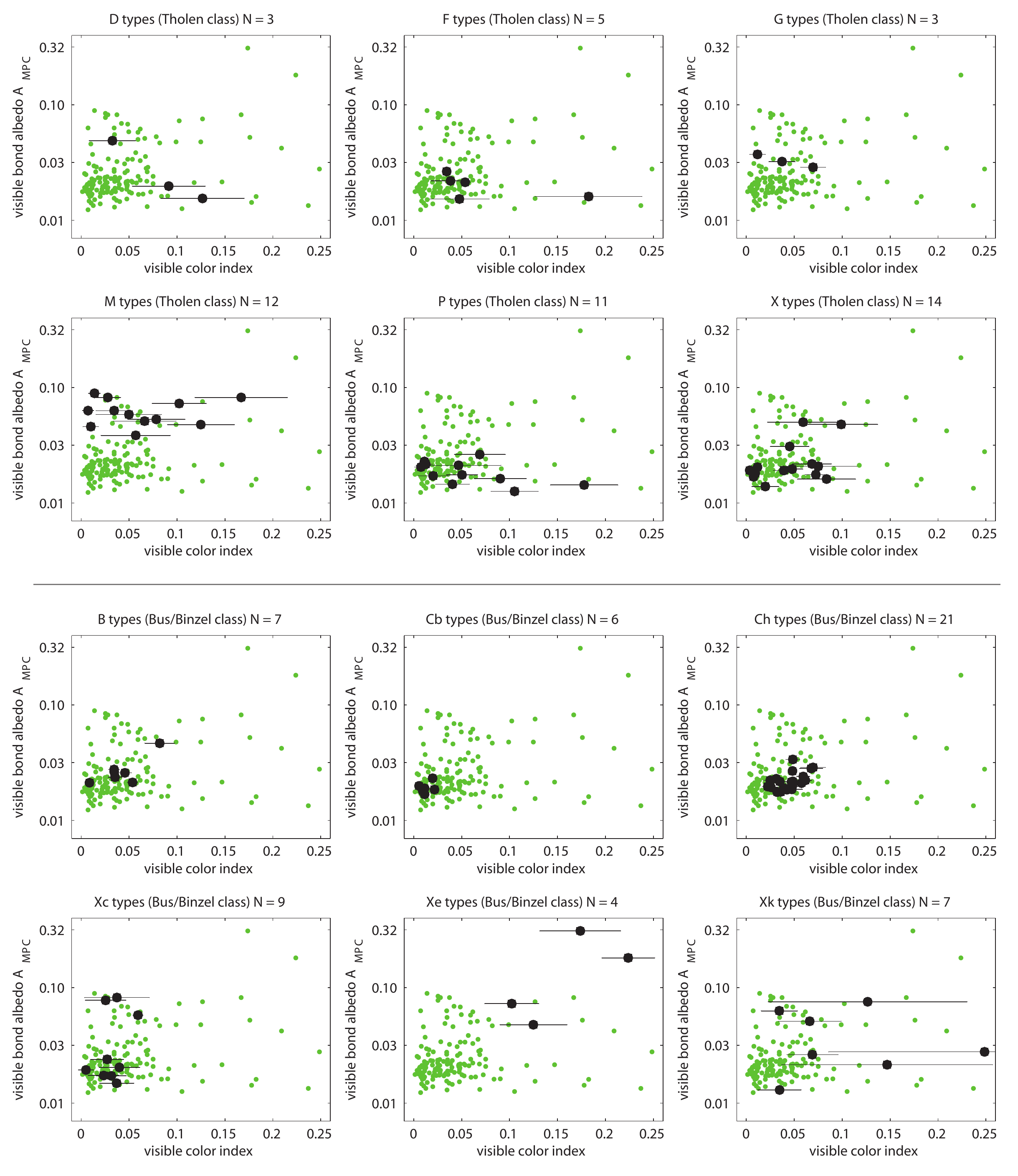}
\caption{Visible color/albedo distributions of Tholen-classified \citep{tho89} and Bus/Binzel-classified \citep{busb02} C-type subgroups among this work's sample of \emph{GALEX}-observed C-type asteroids. See text for further information.}
\vspace{10pt}
\end{figure*}

\begin{figure*}
\centering
\includegraphics[scale=0.74]{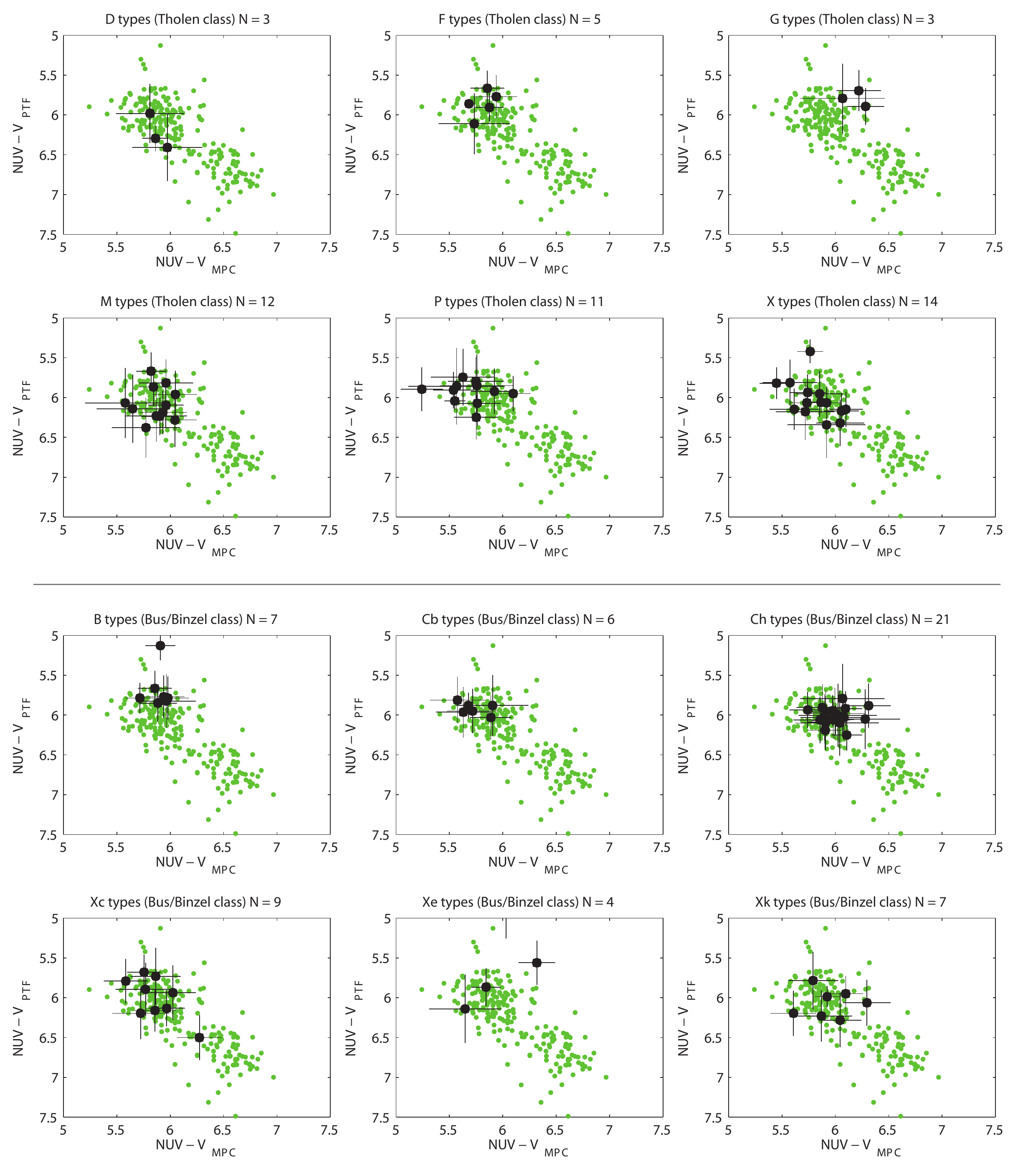}
\caption{$\text{NUV}-V$ color distributions of Tholen-classified \citep{tho89} and Bus/Binzel-classified \citep{busb02} C-type subgroups among this work's sample of \emph{GALEX}-observed C-type asteroids. See text for further information.}
\vspace{10pt}
\end{figure*}

\section{C-type subgroups}

C types deserve further consideration for several reasons: (1) C types outnumber S types in the \emph{GALEX} samples by a factor of several, (2) our division of C types into C$_\text{low}$ and C$_\text{high}$ merits interpretation in more conventional taxonomic systems, and (3) both of the \emph{HST}-observed asteroids in the previous section are known members of C-type subgroups, the UV properties of which are worth confirming with additional group members.

Figures 10 and 11 detail the distribution of \emph{GALEX}-observed asteroids belonging to six classes each from the Tholen and Bus/Binzel taxonomic systems (\citealp{tho89}; \citealp{busb02}), the latter is sometimes referred to as the SMASSII system after the survey data with which it was derived. These two classification systems were created on the basis of different visible-band color data; a comparison of their group definitions is given in Table 1 of \cite{bus02}. We consider only the subset of \emph{GALEX}-observed asteroids having both $V_\text{MPC}$ and $V_\text{PTF}$ model magnitudes and omit subgroups containing less than three objects. In the following subsections we briefly comment on these subgroups. 

One key interpretation of these data---supported also by the \emph{HST} data in Figure 9---is that NUV-band albedo is not very useful for discriminating C-type subgroups, {\it e.g.}, M types versus G types, whereas $u$ band appears to be more diagnostic in this regard. The $u$-band discrepancy between these subgroups was remarked most notably by \cite{zel85} in the ECAS data, but it was unknown at that time (indeed, up until now) whether the discrepancy in UV albedo became more or less pronounced shortward of $\sim$300 nm. The NUV data indicate that the discrepancy lessens in the NUV, as M types do in fact exhibit a step-down in albedo (between NUV and $u$ bands), similar to the step down the G types exhibit within $u$ band.

\subsection{X complex}

The Tholen system's X-type group includes asteroids with relatively flat visible color, including no substantial absorption in the blue (in contrast to, {\it e.g.}, the $u$-band drop-off seen in G types). The subgroups within the X group include E, M, and P types and are distinguishable only by albedo. 

The twelve M types in our sample all have $A_\text{MPC}>0.03$ and $p_{W1}>0.125$, the latter formally makes them all C$_\text{high}$ types in this work's classification system. The M types have $\text{NUV}-V_\text{MPC}=5.89\pm0.15$ and $\text{NUV}-V_\text{PTF}=6.12\pm0.21$, neither of which significantly differ from the C-type averages given in Table 3. This is consistent with the above-noted observation that M-type Lutetia's $\text{NUV}-V$ is similar to that of the \emph{GALEX} C types, despite an obvious difference in $u$-band (Figure 9). Assuming \emph{all} 29 of the C$_\text{high}$ types in the \emph{GALEX} sample are in fact M types, then the C$_\text{high}$ types' slightly higher $\text{NUV}-V_\text{PTF}=6.14\pm0.33$ (compared to $\text{NUV}-V_\text{PTF}=6.03\pm0.22$ for the whole C type group) agrees well with the M types' slightly higher average.

Complementary to the M types, the eleven P types in our sample all have $A_\text{MPC}<0.03$ and $p_{W1}<0.125$, the latter formally makes them all C$_\text{low}$ types. The P types have $\text{NUV}-V_\text{MPC}=5.74\pm0.17$ and $\text{NUV}-V_\text{PTF}=5.91\pm0.13$. These values are less than both models' C-type averages as well as less than the C$_\text{low}$ average, suggesting our C$_\text{low}$ group includes more diverse objects than just P types ({\it e.g.}, the five F types also all have $A_\text{MPC}$ consistent with C$_\text{low}$).

There are 14 \emph{GALEX}-observed asteroids listed simply as X types in the Tholen system (presumably because no visible albedo was available at the time of classification); Figure 10 shows that these are in fact distributed across both the C$_\text{low}$ and C$_\text{high}$ albedo ranges.

In the Bus/Binzel system, the X complex consists of four subgroups: Xc, Xk, X and Xe, these being differentiated by their spectral slope and presence of various absorption features. In the \emph{GALEX} sample the most numerous of these are the Xc types, which have the least red visible color and seem to include both high and low visible albedo members. Both the Xe and Xk types have higher visible color indices (with larger uncertainties in the color). As with the Tholen X types, we see no systematic trends with respect to the NUV properties of these subgroups.

\subsection{G types}

Three \emph{GALEX}-observed asteroids are categorized as G types. Like G-type Ceres, these have intermediate $A_\text{MPC}$ and an above-average $\text{NUV}-V_\text{MPC}=6.22\pm0.11$. In contrast, however, the G-type $\text{NUV}-V_\text{PTF}=5.79\pm0.10$ lies slightly below the C-type average. The reason for this discrepancy is that all three G types in this sample have $p_{W1}<0.125$ and so are formally classed as C$_\text{low}$ objects, as a result their assumed $A_\text{bond}=0.01$ in the computation of $V_\text{PTF}$ may be too low. On the other hand, \cite{osz11} fit $G_{12}=0.88\pm0.2$ to Ceres' phase function, suggesting that the assumed $G_{12}=0.84\pm0.14$ for C$_\text{low}$ types (Table 2) is a more valid assumption for G types than the C$_\text{high}$ value of $G_{12}=0.42\pm0.20$. Hence the G types seem not to fit well into either of our C$_\text{low}$ or C$_\text{high}$ groups, and hence are not accurately modeled by our $V_\text{PTF}$.

The three-asteroid G-type sample's higher than average $\text{NUV}-V_\text{MPC}$ agrees with the Ceres \emph{HST} data (Figure 9), which as discussed above could be indicative of an absorption feature at $\sim$300 nm unique to G types \citep{li06}, the precise shape and location of which remains unresolved in the broadband photometry considered here.

Tholen's G types are represented in the Bus/Binzel system by the Cg and Cgh groups; however no asteroids in our \emph{GALEX} sample have either of these SMASSII labels.

\subsection{B types}

Members of the Tholen B and F classes, represented also by the Bus/Binzel B and Cb classes, all are classified as C$_\text{low}$ types in the \emph{GALEX} sample based on their $p_{W1}$. Unlike the G types, the B types are not anomalous in $\text{NUV}-V_\text{MPC}$, meaning the B types likely lack the G types' strong absorption at 300nm. The B types also are characterized by slightly higher $A_\text{MPC}=0.026$ compared to the C$_\text{low}$ average $A_\text{MPC}=0.020$. Hence, like the G types, the B types show a lower than average $\text{NUV}-V_\text{PTF}$ symptomatic of an underestimated $A_\text{bond}$ and therefore too dim of a predicted $V_\text{PTF}$.

\section{Summary}

We present NUV-band photometry of 405 asteroids observed serendipitously by \emph{GALEX} from 2003--2012. Using a compilation of visible-band color data, we select the subset of these \emph{GALEX}-observed asteroids belonging to the C-type or S-type classes. We then compute the visual-band magnitude (using two different models) corresponding to each \emph{GALEX} detection in an effort to study the $\text{NUV}-V$ color. For both $V$ models, the derived $\text{NUV}-V$ color distribution is bimodal, with S types having the redder color, just as they do within the visible band. The average C-type $\text{NUV}-V$ agrees with \emph{HST} observations of the asteroids Lutetia and Ceres, both of which are members of the visible-color-defined C-type group. Slight differences in the measured $\text{NUV}-V$ among known taxonomic subgroups of the C types may indicate membership in either the M-type or G-type subgroups, though the 300--400 nm region ($u$-band) is more diagnostic of this division.

\begin{table*}
\centering
\caption{Glossary of acronyms and symbols used repeatedly in this work.}
\begin{tabular}{cl}
\hline\\[-2ex]
 symbol or & \multirow{2}{*}{full meaning/description} \\
acronym & \\
\hline\\[-2ex]
$\alpha$ & solar phase angle (angle drawn by a light ray as it travels from the Sun to an asteroid to the Earth)\\
$A_\text{bond}$ & visible-band bond albedo \\
$A_\text{MPC}$ & visible-band bond albedo computed using $H$ and $G$ values from the MPC together with an infrared-derived diameter \\
$A_\text{NUV}$ & near-ultraviolet-band bond albedo \\
C$_\text{high}$ types & C-type asteroids with a high near-infrared albedo ($p_{W1}>0.125$), a group consisting almost exclusively of M types\\
C$_\text{low}$ types & C-type asteroids with a low near-infrared albedo ($p_{W1}<0.125$), a group consisting of P types and many X types\\
$D$ & asteroid diameter\\
ECAS & Eight-Color Asteroid Survey \\
FUV & far-ultraviolet band ($\lambda\lesssim180$ nm)\\
$G$ & an older photometric phase-function model parameter \citep{bow89}\\
$G_{12}$ & a newer photometric phase-function model parameter \citep{mui10}\\
G types & certain asteroids, including Ceres, that are a subgroup of the C-type asteroid taxonomic class\\
\emph{GALEX} & Galaxy Evolution Explorer satellite \\
$H$ & visible-band absolute magnitude ($V$ magnitude asteroid would have if observed 1 AU from both the Sun and Earth, at zero phase angle)\\
\emph{HST} & Hubble Space Telescope\\
\emph{IUE} & International Ultraviolet Explorer satellite\\
MPC & Minor Planet Center, \href{http://minorplanetcenter.net}{\color{blue}{http://minorplanetcenter.net}}\\
 NUV & near-ultraviolet band (180--200 nm), and/or measured magnitude in this band\\
PTF & Palomar Transient Factory survey\\
$p_V$ & visible-band geometric albedo \\
$p_{W1}$ & near-infrared ($W$1-band from \emph{WISE}) geometric albedo\\
$\rho_\text{Spearman}$ & Spearman's correlation coefficient\\
SDSS & Sloan Digital Sky Survey\\
SMASS & Small Main-belt Asteroid Spectroscopic Survey\\
$V$ & visible-band ($\sim$600 nm) astronomical magnitude\\
$V_\text{MPC}$ & predicted $V$ magnitude based on MPC-hosted observational data and the $G$ phase-function model\\
$V_\text{PTF}$ & predicted $V$ magnitude based on color-class-averaged albedos and phase-functions data derived from PTF data and $G_{12}$ phase-function model\\
$V$ & visible-band ($\sim$600 nm) astronomical magnitude\\
\emph{WISE} & Wide-field Infrared Explorer satellite\\
\hline
\end{tabular}
\smallskip
\end{table*}

\section*{Acknowledgements}

A. Waszczak has been supported in part by the W.M. Keck Institute for Space Studies (KISS) at Caltech. E.O.O. is incumbent of the Arye Dissentshik career development chair and is grateful to support by grants from the Willner Family Leadership Institute Ilan Gluzman (Secaucus NJ), Israeli Ministry of Science, Israel Science Foundation, Minerva and the I-CORE Program of the Planning and Budgeting Committee and The Israel Science Foundation.

This work makes use of data products from the {\it Galaxy Evolution Explorer} mission, developed and operated with support from JPL/Caltech, the Centre National d'Etudes Spatiales of France and the Korean Ministry of Science and Technology, Orbital Sciences, the University of California at Berkeley, and the Laboratoire d’Astrophysique de Marseille.

This work also makes use of data derived from the \emph{Palomar Transient Factory} (PTF) and the \emph{Intermediate PTF} (iPTF) projects, operated at the 1.2-m Samuel Oschin Telescope at Palomar Observatory. Participating institutions have included Caltech, Columbia University, Las Cumbres Observatory Global Telescope Network, Lawrence Berkeley National Laboratory, the National Energy Research Scientific Computing Center, the University of Oxford, the Kavli Institute for the Physics and Mathematics of the Universe, Los Alamos National Laboratory, the Oskar Klein Centre, the University System of Taiwan, the University of Wisconsin Milwaukee, and the Weizmann Institute of Science.

This work also makes use of data products from the \emph{Wide-Field Infrared Survey Explorer}, which is a joint project of the University of California Los Angeles and the Jet Propulsion Laboratory (JPL)/Caltech, funded by NASA. This publication also makes use of data products from \emph{NEOWISE}, which is a project of JPL/Caltech, funded by the Planetary Science Division of NASA. 

This work also makes use of data from the Sloan Digital Sky Survey (SDSS), managed by the Astrophysical Research Consortium for the Participating Institutions and funded by the Alfred P. Sloan Foundation, the Participating Institutions, the National Science Foundation, the US Department of Energy, NASA, the Japanese Monbukagakusho, the Max Planck Society, and the Higher Education Council for England.

\section*{Glossary of Acronyms and Symbols}

For the reader's convenience, Table 4 summarizes the various acronyms and mathematical symbols used in this work.

\end{document}